\documentclass[12pt,preprint]{aastex}

\newcommand{\myemail}{rmason@gemini.edu}
\def\deg{\hbox{$^{\:\circ}$~}}
\def\um{$\mu$m}

\shorttitle{Dust and PAHs  in NGC~1097}

\shortauthors{Mason et al.}

\begin{document}
\title{Dust and PAH emission in the star-forming active nucleus of NGC~1097}

\author{R. E. Mason}
\affil{Gemini Observatory, Northern Operations Center, 670, N. A'ohoku Place, Hilo, HI 96720 USA}
\email{\myemail}

\author{N. A. Levenson}
\affil{Department of Physics and Astronomy, University of Kentucky, 177 Chemistry/Physics Building, Lexington, KY 40506, USA}
\email{levenson@pa.uky.edu}

\author{C. Packham}
\affil{Department of Astronomy, University of Florida, PO Box 112055, 211 Bryant Space Center, Gainesville, Fl 32611, USA}
\email{packham@astro.ufl.edu}

\author{M. Elitzur}
\affil{Department of Physics and Astronomy, University of Kentucky, 177 Chemistry/Physics Building, Lexington, KY 40506, USA}
\email{moshe@pa.uky.edu}

\author{J. Radomski}
\affil{Gemini Observatory, Southern Operations Center, Casilla 603, La Serena, Chile}
\email{jradomski@gemini.edu}

\author{A. O. Petric}
\affil{Astrophysics Department, Columbia University, 520 W. 120th Street, New York, NY10025, USA.}
\email{andreea@astro.columbia.edu}

\author{G. S. Wright}
\affil{UK Astronomy Technology Center, Royal Observatory Edinburgh, Blackford Hill, Edinburgh, EH9 3HJ, UK}
\email{gsw@roe.ac.uk}

\begin{abstract}

The nucleus of the nearby galaxy, NGC~1097, is known to host a young, compact (r $<$ 9 parsec) nuclear star cluster as well as a low-luminosity active galactic nucleus (AGN). It has been suggested both that the nuclear stellar cluster is associated with a dusty torus, and that low-luminosity AGN like NGC~1097 do not have the torus predicted by the unified model of AGN. To investigate these contradictory possibilities we have acquired Gemini/T-ReCS 11.7~$\mu$m and 18.3~$\mu$m images of the central few hundred parsecs of this galaxy at $<$45 parsec angular resolution, in which the nucleus and spectacular, kiloparsec-scale star-forming ring are detected in both bands. The small-scale mid-infrared (mid-IR) luminosity implies thermal emission from warm dust close to the central engine of this galaxy. Fitting of torus models shows that the observed mid-IR emission cannot be accounted for by dust heated by the central engine. Rather, the principal source heating the dust in this object is the nuclear star cluster itself, suggesting that the dust that we detect is not the torus of AGN unified schemes (although it is also possible that the dusty starburst itself could provide the obscuration invoked by the unified model). Comparison of {\em Spitzer}/IRS and Gemini/GNIRS spectra shows that, although polyaromatic hydrocarbon emission (PAH) bands are strong in the immediate circumnuclear region of the galaxy, PAH emission is weak or absent in the central 19 parsecs.  The lack of PAH emission can probably be explained largely by destruction/ionization of PAH molecules by hard photons from the nuclear star cluster. If NGC~1097 is typical, PAH emission bands may not be a useful tool with which to find very compact nuclear starbursts even in low-luminosity AGN.

\end{abstract}


\section{Introduction}
\label{intro}

The dusty molecular torus, cornerstone of the unified model of active galactic nuclei (AGN), has been the subject of much observational and theoretical scrutiny in recent years.
The detection of broad emission lines in polarised light in a number of type 2 objects \citep{AM85,Miller90,Young96b,Moran00,Tran01} demonstrates the overall validity of the unified model, and structures thought to correspond to the torus itself have now been resolved through mid-IR interferometry of one Seyfert galaxy \citep{Jaffe}. However, questions remain as to the origin and precise nature of the obscuration and the extent to which it can account for the differences between the many classes of AGN. 

In particular, the possibility that the torus and star formation are intimately connected has been raised by several authors on theoretical grounds. For instance, the material in the nucleus needs to have a certain scale height to obscure the nucleus and account for the relative numbers of types 1 and 2 AGN, and it has been suggested that the energy necessary to maintain this thickness might be supplied by supernovae in nuclear starbursts  \citep{Wada02}. Indeed, some or all of the nuclear obscuration has been proposed to arise in dusty clouds ejected from such starbursts \citep{CidFernandes95,Fabian98,Watabe05} or in optically thick stellar winds \citep{Zier02}. Observationally, young stars are found quite commonly in the inner few hundred pc of AGN \citep[e.g.][]{Gonzalez98,StorchiBergmann01,Imanishi02,Imanishi03,Rodriguez03}.

However, while few-hundred-parsec-scale star formation is often observed in AGN,  few young stellar clusters have to date been detected at smaller scales. The outer diameter of the torus remains rather unclear, but its warm, mid-IR-emitting regions have not yet been resolved in single-dish 10-20 \um\ observations of even the nearest objects \citep[e.g.][]{Bock2,Radomski03,Soifer03,Packham05,Mason06}, and interferometric measurements of NGC~1068 suggest that its mid-IR-emitting regions are only $\sim$3 pc ($\sim$0.04\arcsec) in diameter \citep{Jaffe}. Near-IR adaptive optics imaging of the Circinus galaxy \citep{Prieto04b} resolves a source of FWHM$\sim$2 pc, consistent with a parsec-scale torus at those wavelengths.

The discovery by  \citet[][hereafter SB05]{StorchiBergmann05} of UV absorption signatures of a 10$^{6
}$M$_{\sun}$, 10$^{6}$ year-old stellar cluster within only 9 pc of the nucleus of the nearby LINER/Seyfert 1 galaxy, NGC 1097\footnote[1]{Although NGC 1097 hosts a classic LINER nucleus \citep{Phillips84}, broad, double-peaked emission lines in its spectrum are characteristic of rotating gas in an AGN accretion disk \citep{StorchiBergmann93}. }, is therefore of obvious interest. The star cluster --- but not the broad emission lines from the nucleus itself --- suffers about 3 magnitudes of visual extinction, consistent with the possibility of starburst activity physically associated with a dusty AGN torus.  Young superstellar clusters or recent star formation are also known to exist in the central few parsecs of a handful of other AGN \citep{Colina02,Davies06,MuellerSanchez06}. 

While the results of SB05 are suggestive of a starburst genuinely associated with an obscuring torus, the observation that many LINERs host nuclear UV point sources raises some doubt as to whether LINER-type nuclei in fact host obscuring tori \citep{Maoz05}. This is expected on theoretical grounds in the context of the disk-wind scenario for the torus: below a certain luminosity the accretion onto the black hole can no longer sustain the mass outflow necessary to obscure the nucleus \citep{Elitzur06}. Moreover, how the properties of LINER tori, if they exist, might relate to those observed in Seyfert galaxies is completely open to question. Further observations of LINER nuclei to probe their detailed nuclear structure are therefore desirable. 

In addition, the UV evidence of star formation detected by SB05 is very susceptible to extinction, so UV detections of starbursts in AGN tori are likely to be limited to cases of nuclei in which the stellar clusters are not too deeply buried within the nuclear dust. Interpreting the energetics of UV-detected starbursts will  be complicated by uncertain extinction corrections. A more useful tracer of the nuclear starburst phenomenon would be an infrared star-formation indicator that is both unaffected by dust extinction and accessible from the ground, to take advantage of the  spatial resolution achievable from ground-based telescopes. The 3.3 \um\ polycyclic aromatic hydrocarbon (PAH) emission band fits both of these criteria. Usually attributed to vibrational relaxation of aromatic hydrocarbon molecules excited by far-UV photons, the PAH bands have been used extensively as tracers of star formation in extragalactic environments \citep[e.g.][]{Moorwood86,Roche91,Genzel98,Lutz98,Rigopoulou99,Laurent00,Tran01,Imanishi04,Peeters04,Risaliti06}.

The suite of PAH emission features is commonly detected in large-aperture 
ISO spectra of nearby active galaxies with circumnuclear star-forming regions \citep{Rigopoulou99}, but on moving to apertures of 3-5\arcsec\ ($\sim$ few hundred pc) or less, the bands are often weak or absent \citep{Roche91,Lefloch,Sieb04,Roche06a,Roche06b}. Modelling suggests that the carriers of the bands are easily destroyed by the harsh radiation field close to the active nucleus itself \citep{Roche85,Sieb04,Voit92}. However, this appears not always to be the case. For instance, \citet{Imanishi04} find that the surface brightness of the nuclear PAH emission in Seyfert 1 galaxies is comparable to that in starburst galaxies, and the 3.3~\um\ PAH band has been shown to be present in several Seyfert galaxies of both types 1 and 2 on fairly small spatial scales, including, in the case of NGC~3227, within only 60 pc of the central engine \citep{Rodriguez03}. To exist at such small distances from these AGN, it seems likely that the fragile PAH band carriers are shielded from the nuclear X-rays. This intriguing result could perhaps be explained if the carriers exist within the dense, dusty clouds of the torus itself.

To further investigate this possibility, we have performed L band ($\sim$3--4~\um) spectroscopy of the nucleus of NGC~1097 (D=17 Mpc for $H_{0}=75$ km sec$^{-1}$ Mpc$^{-1}$; 1\arcsec = 85 pc) to 
search for the 3.3~\um\ PAH emission band. We have also obtained images at 11.7 and 18.5~\um, to search for the compact mid-IR source expected to be a signature of the torus and to constrain its properties.  The data were taken using the Thermal Region Camera and Spectrograph (T-ReCS)  and the Gemini Near-Infrared Spectrograph (GNIRS) on the Gemini South 8.1 m telescope, yielding $\le$0.52\arcsec\ spatial resolution and therefore probing distances $\le$ 22 pc from the nucleus.
We also discuss complementary {\em Spitzer Space Telescope} Infrared Spectrograph (IRS) observations of the central 3.6 $\times$ 3.6\arcsec\   (approx. 300 $\times$ 300 pc) of this galaxy. \S \ref{obs} describes all of these observations and the reduction of the data. In \S \ref{results} we present the nuclear data and describe the characteristics of the compact source detected in the mid-IR, and compare our subarcsecond-resolution data and that of \citet{Prieto05} with the clumpy torus models of \citet{Nenkova}.  The conclusions that can be drawn from this work are discussed in \S \ref{discuss}.


\section{Observations and data reduction}
\label{obs}

\subsection{Mid-IR imaging}

Narrow-band N (11.7~$\mu$m) and Q (18.3~$\mu$m) band images of NGC~1097 were taken on 2005 September 16 and 17 using
the Gemini South mid-IR instrument T-ReCS \citep{Telesco98}.
T-ReCS uses a Raytheon 320 x 240 pixel Si:As IBC
array, providing a plate scale of 0.089$\arcsec$ per pixel.
The detector was used in correlated quadruple
sampling mode \citep{Sako03}.  Images were obtained in the
11.7$\mu$m ($\delta$$\lambda$ = 1.13$\mu$m,
50\% cut-on/off) and 18.3$\mu$m ($\delta$$\lambda$ = 1.51$\mu$m,
50\% cut-on/off) filters using
the standard chop-nod technique to remove time-variable sky
background, telescope thermal emission and "1/f" noise.  The chop throw was 15\arcsec\ and the telescope
was nodded 15\arcsec\ in the direction of the chop every 30 s.  The chop angle was chosen based on the 5\arcsec-aperture N band data of \citet{Telesco81} to minimise the extended emission in the off-source beam and was fixed at
90\deg east of north. The total on-source time for the 11.7~$\mu$m and 18.3~$\mu$m observations was 456 s and 912 s respectively

The data were reduced using the Gemini IRAF package. The
difference for each chopped pair for each nod-set was
calculated, and the nod-sets then differenced and combined until a
single image was created.  Chopped pairs obviously compromised 
by variable sky background, high electronic noise, or other problems were searched for but no data needed to be discarded. 

The Cohen standards SAO~216405 
and SAO~193679,
and PPM~278407 were observed for flux and PSF calibration.
Observations of the standards taken approximately 1 hr apart 
showed a variation in counts of around 10\%, typical of
mid-IR photometry.  The PSF observations
were made immediately before or after the NGC~1097 observations
and with the same instrumental configuration.  The  FWHM of the PSF star  was 0.41\arcsec\ 
(standard deviation $\sim$0.03\arcsec) at 11.7 $\mu$m and 
0.52\arcsec\ in the 18.3 $\mu$m filter. Short PSF or flux standard
observations are comparable to longer
source observations as the closed-loop active optics
of Gemini provides a similar PSF when taken at a
similar telescope pointing and time \citep[see e.g.][although seeing variations may of course affect the final observed PSF]{Radomski03,Packham05}.  As judged by the FWHM of the standard stars, the PSF was stable over the course of these observations.

\subsection{L-band spectroscopy}

L band spectroscopy of the nucleus of NGC~1097 was obtained using GNIRS, the near-IR spectrograph  on the Gemini South telescope, on 2005 November 13.  The short red camera and 3-pixel (0.45\arcsec) slit were used together with the 32 l/mm grating, resulting in R$\sim$1100.  The total integration time was 30 minutes, and the telescope was nodded by 20\arcsec\  every 30 seconds in an ABBA pattern so that the sky background emission could be subtracted out. Sky conditions were photometric during these observations.

The nucleus was acquired in the H band to take advantage of the contrast between object and sky at those wavelengths. A point source contributes 52\% of the H band flux in a 0.4\arcsec\ diameter aperture, that fraction increases with wavelength \citep{Prieto05}, and we observe an unresolved source at 11.7~\um\  (\S~\ref{images}), suggesting that the location at which the slit was centered at 1.6~$\mu$m also corresponds to the flux peaks at 3.3 and 11.7~\um. The slit was orientated at 9\deg east of north in order to encompass one of the circumnuclear emission knots visible in the T-ReCS image, but that knot proved too faint to yield a useful spectrum and will not be considered further.

The data were reduced using the Gemini IRAF package in combination with standard IRAF and Starlink FIGARO routines. After bad pixel masking, flatfielding and subtracting in pairs, the individual sky-subtracted spectra were averaged together, with the positive and negative beams cross-correlated and shifted in both the spatial and dispersion directions before combining. The central three pixels (0.45\arcsec) of the galaxy spectrum
were extracted using the standard star spectrum to define the curvature of the extraction aperture across the array. 

Wavelength calibration was accomplished using the numerous telluric absorption lines in the spectrum. 
The galaxy spectrum was then cross-correlated with the spectrum of the standard star, BS 466 (F7V),  and divided by that spectrum. In order to improve the cancellation of telluric lines, the resulting spectrum was multiplied by a logarithmically-scaled model atmospheric spectrum\footnote[2]{\citet{Lord92}, obtained from www.gemini.edu/sciops/ObsProcess/obsConstraints/ocTransSpectra.html}, binned to the appropriate spectral resolution. This had the effect of slightly reducing the amplitude of some of the residuals in the 3.3~$\mu$m region. The spectrum was then multiplied by a 6240 K blackbody curve. Flux calibration was with reference to BS 466, using a V band magnitude of 6.24 and V-L = 1.36. No attempt was made to correct for uncertain differential slit losses between the galaxy and standard star (due to e.g. seeing variations and different intrinsic spatial flux distributions) and, as is usual for flux calibration of narrow-slit spectral data,  the uncertainty of the derived flux density is considerable.

\subsection{IRS spectroscopy}

Publically-available spectra of the nucleus of NGC~1097 taken with the short-low (first order) module of the {\em Spitzer Space Telescope}'s Infrared Spectrograph \citep[][IRS]{Houck04} were retrieved from the Spitzer data archive (PID 14; PI Houck).  These R$\sim$100 observations represent a total integration time of 56 s.  Files output from the IRS pipeline at the Spitzer Science Center (which includes ramp fitting, dark-sky subtraction, droop correction, linearity correction, flatfielding and wavelength calibration) had background light subtracted  by subtracting the first-order slit exposed during second-order observations of NGC~1097 taken a few minutes previously. Spectra were extracted from the background-subtracted, coadded frames using Spice, the SPItzer Custom Extractor. The extraction window scales with wavelength but was defined to be 2 pixels (3.6\arcsec) at 11.2~$\mu$m, equal to the 3.6\arcsec\ slit width and comparable to {\em Spitzer}'s diffraction limit at 11.2 \um\ (3.3\arcsec).

The spatial profile of the IRS spectrum shows flux between the nucleus and the circumnuclear ring $\sim$10\arcsec\ distant, presumably from the wings of the PSFs of these bright sources combined with real diffuse emission from that region. Neither the point source nor the extended source flux calibration provided by Spice is optimised to deal with this situation. However, the conclusions drawn from these observations do not depend critically on a highly accurate calibration, so we simply normalise the mean flux in the IRS spectrum to that given by \citet{Telesco81} over a similar bandpass in a slightly larger aperture (5\arcsec\ diameter vs. 3.6 $\times$ 3.6\arcsec).


\section{Results}
\label{results}

\subsection{The mid-IR images and model fits}
\label{images}

\begin{figure}[t]
\includegraphics[angle=-90,width=12cm]{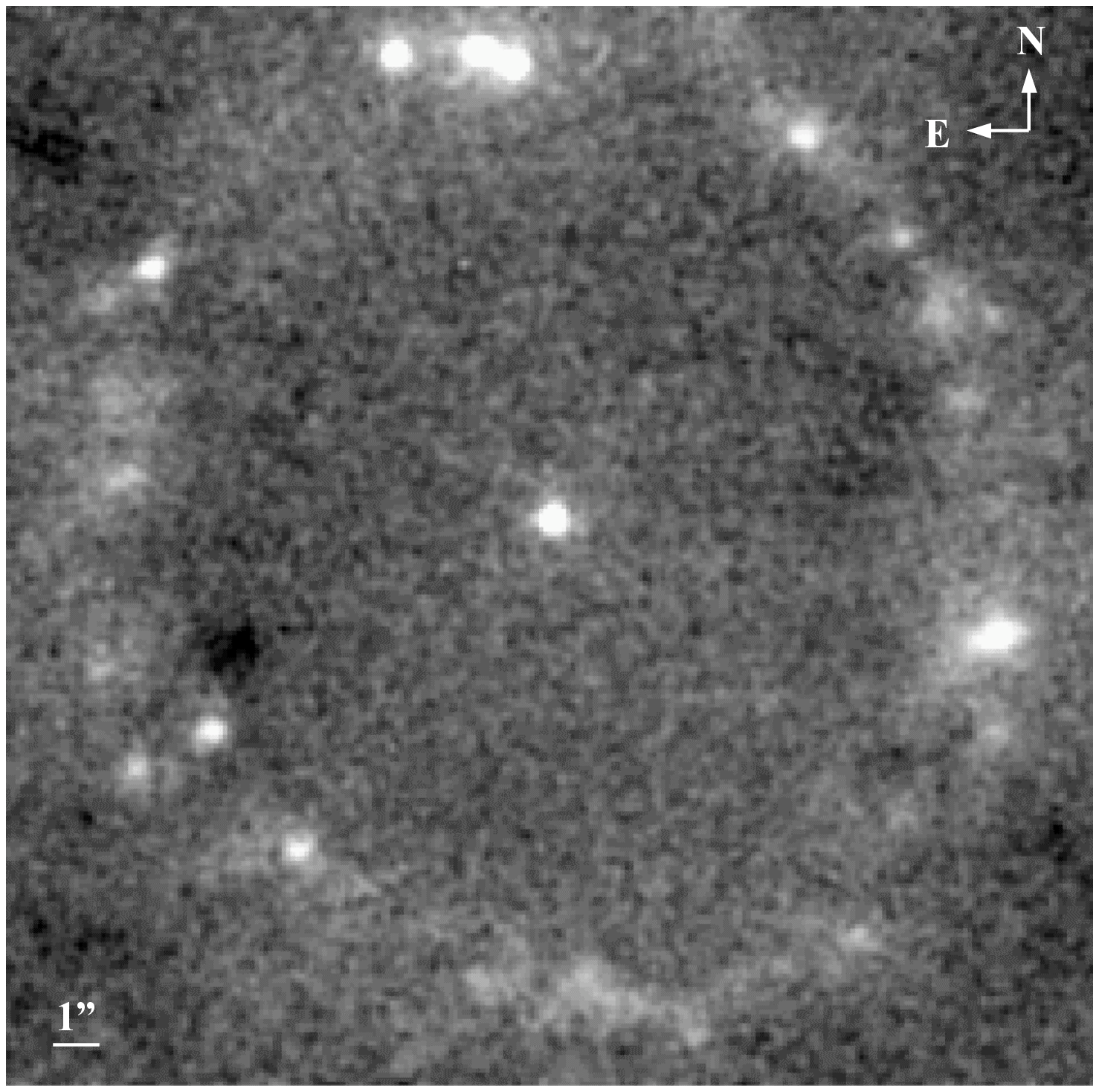}
\caption{Central portion of the T-ReCS 11.7 image of NGC1097, smoothed with a 2-pixel Gaussian.  The (unguided) negative off-source images are evident at either side of the positive galaxy image.}
\label{fig:Si5}
\end{figure}

\begin{figure}[t]
\includegraphics[angle=0,width=12cm]{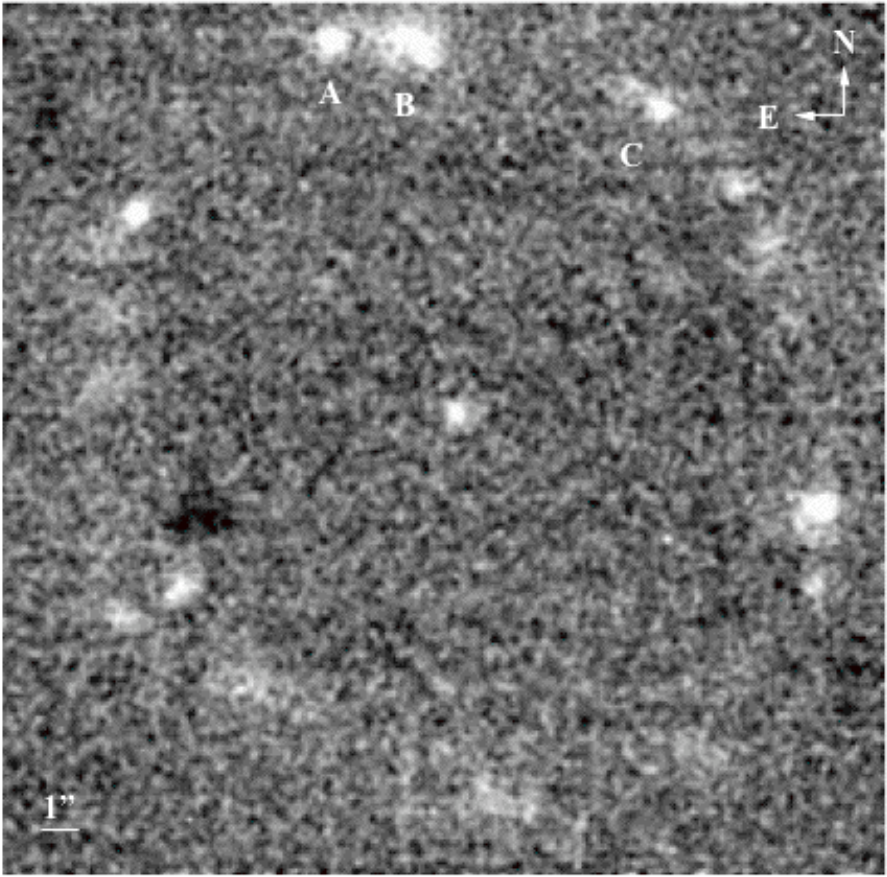}
\caption{As for Figure 1, but at 18.3 $\mu$m. Photometry of clumps A, B and C is given in Table~\ref{table1}.}
\label{fig:Qa}
\end{figure}

The T-ReCS mid-IR images of NGC~1097 are shown in Figures~\ref{fig:Si5} and ~\ref{fig:Qa}. A compact nuclear source and the well-known circumnuclear star-forming ring are clearly detected at both 11.7 and 18.3 $\mu$m. The starburst ring emits strongly in the mid-IR, with each of the three knots in the northern part of the ring 
contributing a similar level of flux to that of the nucleus itself.  The 15\arcsec\ maximum chop throw of the Gemini telescopes means that the positive and negative images partially overlap in the region of the circumnuclear ring. Inside the ring, faint extended emission is very probably present but is difficult to detect without a clear sample
of the sky on which to measure the zero level of the background.  The flux of the nuclear point source is approximately 48 mJy at 11.7 $\mu$m and 64 mJy at 18.3 ~$\mu$m (Table~\ref{table1}), roughly 75\% and 27\% of the broad-band, 5\arcsec-aperture measurements of \citet{Telesco81} at N and Q respectively, and consistent with the 60 mJy, 1.5\arcsec-aperture  N-band upper limit of \citet{Gorjian04}. Faint emission between the nucleus and ring likely accounts for the difference between our data and that of \citet{Telesco81}, most clearly affecting the Q band photometry.

Table~\ref{table1} gives the flux densities of the three regions of emission in the circumnuclear ring that are clearly detected at 18.3~$\mu$m and least likely to be affected by the negative chop beams. The aperture sizes used were determined partly by the proximity of neighboring emission regions and/or the edge of the detector. The N/Q flux density ratios are in the region of 0.4 for the ring, but rather higher, about 0.75, for the nucleus.  

\begin{deluxetable}{cccc}
\tabletypesize{\scriptsize}

\tablecaption{Photometry of the nucleus and circumnuclear ring of NGC~1097 }

\tablewidth{0pt}

\tablehead{
\colhead{Location\tablenotemark{1}} & \colhead{Aperture diameter} & \colhead{11.7~$\mu$m flux density} & \colhead{18.3~$\mu$m flux density} \\
\colhead{} & \colhead{(arcsec)} & \colhead{(mJy)} & \colhead {(mJy)}
}
\startdata
 Nucleus & 3.0\tablenotemark{2} & 48   & 64   \\
 A & 1.6 & 34   & 83   \\
 B & 2.0 & 75   & 169 \\
 C & 1.6 & 25   & 63   \\
  \enddata
\tablenotetext{1}{See Fig.~\ref{fig:Qa}}
\tablenotetext{2}{An aperture of 3" diameter was used for the photometry of the  nucleus, chosen based on the profiles of the standard stars to ensure  that all of the flux from the unresolved source was collected. A  surrounding "sky" annulus was used to subtract off any residual sky  background and diffuse emission surrounding the nucleus.  The  (unknown) spatial variation of surrounding diffuse emission will therefore determine the degree to which the photometry represents the unresolved source alone,  but as any such emission must be close to the sensitivity limit of the instrument, we expect it to make a negligible contribution to the measured nuclear flux.}

\label{table1}
\end{deluxetable}

The FWHM of the central compact source was measured to be 0.43\arcsec\ at 11.7~$\mu$m, comparable to the mean FWHM of the comparison star observations  (0.41$\pm$0.03\arcsec, \S~\ref{obs}). This translates to a diameter of $<37$~pc, consistent with the NIR limit of 10~pc diameter of \citet{Prieto05}. Thus, like all AGN observed non-interferometrically to date, the nuclear source of NGC~1097 is not resolved at its FWHM in the N band.  Limited S/N prevented a meaningful measurement of the source size at 18.3~$\mu$m. 

\clearpage

\clearpage

While AGN- or star-formation-heated dust is expected to emit in the mid-IR, other processes can also contribute to the nuclear mid-IR emission. For instance, as shown by \citet{Perlman01} and \citet{Whysong04}, the 10~$\mu$m emission from M87 on subarcsecond scales can be fully accounted for by synchrotron radiation from the inner regions of the radio jet.  However, the thin disk + jet + radiatively inefficient accretion flow (RIAF) model of 
\citet{Nemmen06}, 
which does not include any thermal dust emission, predicts mid-IR luminosity around an order of magnitude lower in both the N and Q bands than that observed from the unresolved nuclear source in NGC~1097 ($\log[\nu L_{\nu}(\rm erg \; s^{-1})]\approx 41.6$ at both N and Q for D=17 Mpc.
Altering the mass accretion rate or other parameters of the thin disk + RIAF 
model cannot account for the observed mid-IR flux, given the observational
constraints at shorter wavelengths.
Even allowing for the fact that the jet parameters in the Nemmen et al. model are not well constrained, 
we conclude 
that the bulk of the mid-IR emission that we detect comes from warm dust in the region of the central engine.

To investigate the contribution of the AGN to the heating of the
nuclear dust, we fit the observed fluxes with the inhomogeneous
torus models of Nenkova et al. (2002).  In these computations,
individual clouds are optically thick, and the inner radius of their
toroidal distribution is located at the dust sublimation radius. 
While the total extent of the torus may reach 30 to 100 times the inner
radius, the smallest physical scales, where the clouds are hottest,
dominate the emission at near/mid-IR wavelengths.  The physical
parameters that describe the distribution of clumps include the average number
of clouds along radial rays in the equatorial plane, $N_0$, 
and the optical depth per
cloud in the $V$ band, $\tau_V$.  The clouds follow a power law
distribution in radius, $r$, with number declining as $r^{-q}$.  The
mid-IR emission is not sensitive to the outer radius of the cloud
distribution, so we fix it at 30 times the inner radius.

We require that the clump distribution give a high probability ($> 0.5$)
of
direct views of the AGN, given the lack of reddening of the broad
H$\alpha$ line and the low X-ray column density toward the nucleus
\citep{Terashima02,Nemmen06}.  \citet{StorchiBergmann03} find the
inclination angle of the accretion disk $i = 34\degr$
\footnote[3]{Note that contrary to the description in their paper,
\citet{StorchiBergmann03} intend $i$ to have its
usual meaning, measured with respect to the rotation axis (not the
plane) of the disk, so edge-on views have $i=90\degr$ (M. Eracleous,
private communication).}.   
We assume that the torus is aligned with the
accretion disk and adopt $i=30\degr$ in the coarse model grid.  The
torus does not have a sharp edge, but rather a Gaussian angular
distribution away from the equatorial plane of width $\sigma$.  
In terms of the angle from the equatorial plane, $\beta$, 
the number of clouds along a ray $N(\beta) = N_0 \exp(-\beta^2/\sigma^2)$.
To
allow direct views to the nucleus with the viewing angle constrained,
we consider $\sigma = 30\degr$, which roughly corresponds to a torus
half-opening angle of $60\degr$, and $\sigma = 45\degr$ 
(for half-opening angle $\sim 45\degr$).
With the broader distribution,
the probability of viewing the unobscured nucleus is generally low,
and only $N_0 \le 2$ is acceptable at $i = 30\degr$.

\begin{figure}[htb]
\includegraphics[angle=0,width=14cm]{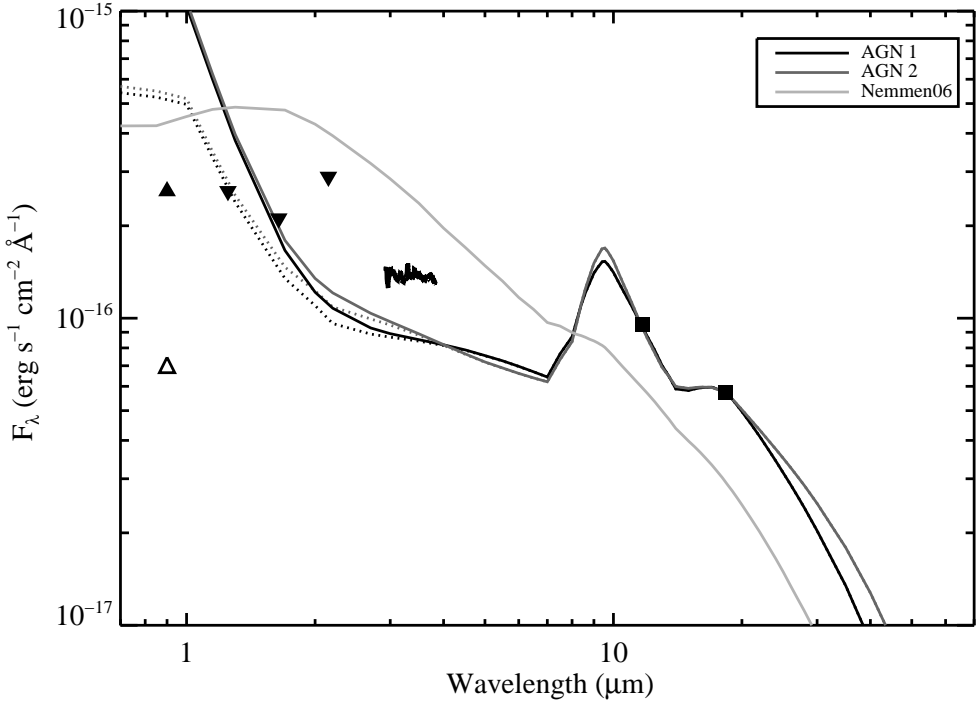}
\caption{Inhomogeneous dusty torus models, fit only to the 11.7  and 18.3~$\mu$m
fluxes ({\it squares}) and constrained to allow a direct view of the
central engine.  
The sum of the intrinsic heating spectrum and the resulting dust
emission is plotted as a  solid line in each case.
Several different obscuring geometries
could produce the mid-IR emission if 
a standard AGN spectrum (\textit{black and dark gray lines})
heats the dusty clouds.  These AGN models 
are scaled to the 18.3~$\mu$m flux, and
applying foreground extinction of the intrinsic source of 
$A_V = 0.12$ mag 
results in the corresponding dotted lines. 
Alternatively, using the fit of \citet{Nemmen06}  as the heating spectrum, none of the allowed
torus geometries gives a good fit to the mid-IR flux ratio or the observed flux in either band.
We plot the best of these
(\textit{light gray line}), fixed to the 
intrinsic luminosity determined by
\citet{Nemmen06}.
None of these AGN or RIAF 
solutions remains consistent
with the observations at all  wavelengths.
We plot a smoothed version of the L-band
spectrum ({\it thick line}) and small-scale near-IR ``AGN magnitudes''
({\it inverted triangles};  see text).  
We also show the total observed emission 
({\it filled triangle}) and reddened starburst contribution
({\it open triangle}) that 
SB05 measure at 9000\AA.
ISO and IRAS data, which could in principle constrain the models in
the far-IR, are not shown as the large apertures involved contain a
very large contribution from emission from the bright, kiloparsec-scale
starforming ring.  See the text
for a complete description of the different models.}
\label{fig:modelagn}
\end{figure}

\begin{figure}[htb]
\includegraphics[angle=0,width=14cm]{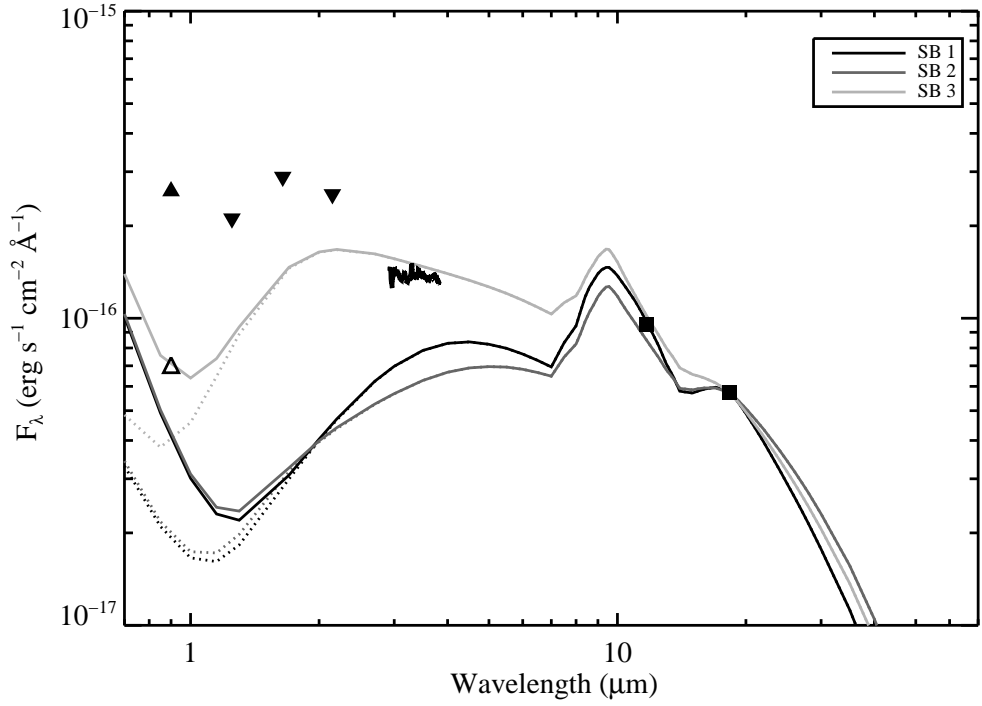}
\caption{Fits of the starburst-heated dust model to the SED of NGC~1097. 
{Solid lines} show the sum of the intrinsic spectrum and
the dust-reprocessed emission.
{Dotted lines} show extinction of the starburst
spectrum by $A_V =3$.
The starburst emission peaks in the UV, so it heats the dust effectively
while
not contributing directly to the net IR flux.  
Data as for Fig.~\ref{fig:modelagn}, and model details are listed in the text.}
\label{fig:modelsb}
\end{figure}

We consider three different heating spectra and
fit the sum of the intrinsic spectrum and the resulting
dusty emission  to the ratio of our N- and Q-band fluxes.
We  scale the
results to the flux at Q  (Figs. \ref{fig:modelagn} and
\ref{fig:modelsb}), 
except when using the result of \citet{Nemmen06}
for the input spectrum, in which case 
multi-wavelength observations already constrain
the luminosity.
Also shown in Figures \ref{fig:modelagn} and \ref{fig:modelsb} are ``AGN magnitudes,'' the J, H and K band flux densities of the nuclear point source derived by fitting unresolved emission and an underlying galaxy component to near-IR imaging data \citep{Prieto05}\footnote[4]{The values given in \citet{Prieto05} are erroneous; corrected values were supplied by M. A. Prieto (private communication).}. We speculate that NGC~1097 will remain unresolved in the mid-IR at resolutions similar to the PSF used by \citet{Prieto05} to fit the near-IR data. Such a result would imply that the mid-IR emitting region is $\la$10 pc in radius, as has been found for other nearby AGN \citep[e.g.][]{Soifer03,Jaffe,Packham05,Roche06b}. We therefore believe it likely that these near-IR data sample the same spatial region as our mid-IR points.

The first dust-heating spectrum we consider is
that of 
a standard AGN (in which $\lambda F_\lambda
\propto \lambda^{-0.5}$ from 0.1 to 1\micron).  
Although the low observed optical and UV fluxes 
together with the minimal obscuration along the line of sight
rule out this model for NGC 1097, it serves as a useful 
point of comparison. 
In this case, the 
dusty material
dominates the mid-IR emission, even when the AGN is viewed directly,
and the AGN itself dominates at
optical and near-IR wavelengths. 
The dust sublimation 
radius scales with the square root of the bolometric luminosity
of the heating source and is a function of the heating spectrum.
This AGN heating spectrum and  the observed 
$L_{bol}= 10^{42} {\rm \; erg \; s^{-1}}$ set 
the inner radius  at 0.01 pc. 
The observed N/Q ratio and a likely unimpeded view
of the central engine exclude $q=1$, 2, and 3 radial distributions.
Two reasonable models both have $q = 0$, $\sigma =30\degr$, and
$N_0 = 2$, with $\tau_V = 40$ and 100 in models ``AGN 1''
and ``AGN 2,'' respectively.
Scaling to the observed flux at Q yields 
$L_{bol} = 4\times 10^{42} {\rm \; erg \; s^{-1}}$, which is a few times
higher than the observed  AGN luminosity.  More severely,
they significantly exceed the observed optical  emission, 
even after correcting for the measured extinction.
This implies that
although dust heating by a standard AGN radiation field can
reproduce the mid-IR SED of NGC~1097  well, this cannot in fact be the
dominant dust-heating mechanism in this galaxy.

\clearpage
\clearpage

\citet{Nemmen06} present the broad-band SED of the nuclear region
of NGC 1097.  The data reveal the intrinsic spectrum at optical and UV
wavelengths, which is responsible for heating the surrounding dust,
although they do not constrain it well in the IR.  Motivated by the
empirical result at short wavelengths, we consider Nemmen et al.'s
thin disk + jet + RIAF model fit to the data as an alternate heating
source of the nuclear dust.  We use this spectral model (which
we refer to as the "Nemmen06" model for simplicity) as the input
to the clumpy torus distribution of clouds.

The Nemmen06 spectrum is weaker than 
the standard AGN at wavelengths below 1$\mu$m 
and relatively stronger than the standard AGN at longer wavelengths.
The  UV-weak continuum heats the clouds inefficiently,
so the dust sublimation temperature is reached closer to the
central source than in the standard AGN model of the same
luminosity.
With the Nemmen06 heating and $L_{bol}= 10^{42} {\rm \; erg \; s^{-1}}$, 
the inner radius is located
at 0.003 pc, while still scaling with the square root of luminosity.
Fixing the viewing angle ($i = 30\degr$), 
none of the $\sigma=30\degr$ models fits the data.
Several of the $\sigma=45\degr$ models with $q =0 $ or 1
agree with the mid-IR flux ratio.  However, 
the required direct view to the AGN rules out these models,
which have $N_0 = 20$ so the probability of an unobscured view
of the nucleus is only 0.03.
For comparison with the standard AGN results, we show the best of the
allowed models based on the Nemmen06 spectrum, in which 
 $q = 0$, $\sigma =45\degr$,
$N_0 = 2$, and $\tau_V = 60$. 
We adopt the observed luminosity of the input Nemmen06 spectrum
and therefore do not scale the result to the mid-IR data as
we did with the standard AGN model.
The resulting model, including the dust emission, 
underestimates the flux density at Q  
by a factor of two, while overestimating the flux density at L 
by a factor of two.

Both the standard AGN  and Nemmen06  spectra produce similar SEDs of
the torus alone, with a peak in $F_\lambda$ around
20--40\um.  
However, because 
the total model is the 
sum of the input and torus spectra,  the significant 
differences between the input spectra
strongly affect the 
relative mid-IR contribution of the torus.
The intrinsic AGN spectrum is very weak in the
mid-IR, so the total mid-IR emission of this model is 
essentially that of the torus alone. 
Here the torus
is thousands of times brighter than the input spectrum
in the mid-IR, even when the torus scale height is small.
In contrast, the Nemmen06 continuum is strong at near- and mid-IR
wavelengths, and its torus does not provide
more than 3 times the flux input at 11.7 or 18.3\um.
Consequently, these models generally overpredict the
observed N/Q flux ratio.
In the Nemmen06 case, the entire resulting optical to mid-IR
spectrum is very sensitive to the input spectrum.
The original model was not constrained well in the IR,
and even without dust reprocessing it overpredicts
the observed near-IR nuclear emission.
A more severe problem of the Nemmen06-heated models is that the 
optical/mid-IR flux ratio is always too high.
The optical/mid-IR signature of the
intrinsic spectrum persists here, 
even when the  dust-reprocessed contribution is included. 
Fundamentally, because
the short-wavelength flux is 
relatively weak, it cannot
produce the strong mid-IR emission we measure.
Thus, we conclude that the complete Nemmen06 model does not
describe the IR data well.
Changing the input contribution of the 
RIAF spectrum alone and neglecting dust-reprocessed emission is not a
viable solution.  The family of RIAF models cannot produce the
observed mid-IR flux density, even when the underlying physical
parameters of the models are varied.

That the clumpy torus models with neither a standard AGN spectrum nor
the Nemmen06 spectrum are able to simultaneously reproduce the mid-IR and the
near-IR/optical emission of NGC~1097 suggests that the nuclear starburst
contributes to the intrinsic luminosity that the dusty clouds
reprocess. In fact, a $10^6 \rm M_\sun$, $10^6$ year-old starburst
scaled by 0.85 (SB05) has $L_{bol} = 5\times10^{42} {\rm \; erg \;
s^{-1}}$ \citep{Leitherer99}, several times higher than the AGN. To
test this possibility we have adapted the clumpy torus models to use
the young starburst spectrum for their input source, as an
approximation to the mixed regions of dust and young stars presumably
present in the nucleus.

We use a spherically-symmetric distribution of clouds with $N_0$ fixed at 0.7, for P$>0.5$ of a direct view of the central engine. The observed N/Q ratio rules out $q=2$ and 3 solutions, but we find good solutions with $q=0$ -- 1 and $\tau_{V}=20$ -- 60:  $q=0$ and $\tau_V = 20$ (``SB 1'' model), $q=0$ and $\tau_V = 40$ (``SB 2'' model), and $q=1$ and $\tau_V = 60$ (``SB 3'' model).
These three models have $L = 1.2$, 1.2, and 
$1.6\times 10^{42} {\rm \; erg \; s^{-1}}$, respectively,
when the model output is scaled to the observed mid-IR flux.
This is
somewhat lower than the starburst luminosity of SB05 but we note that
that value relies heavily on the large extinction correction of the UV
spectrum.  Because we fit only the mid-IR data, changing the
extinction would not alter our preferred model fits, although
it would change the modeled observable optical emission
(dotted lines in Figure \ref{fig:modelsb}).
The starburst heats the dust effectively because of its
high UV luminosity, but its weakness in the infrared relative to both
the standard AGN power-law and RIAF spectra
means that the near-IR emission is predominantly from
the dusty clouds, rather than from a bright AGN component that
overpredicts the emission at optical wavelengths. 
The geometry
of the starburst models is somewhat artificial, and we have no
direct measurement of the spatial relationship between
the starburst-heated clouds and the accretion disk.
All the flux measurements include both the starburst
and the central engine along with the emission they 
reprocess on  scales of 10 pc.
Nevertheless, 
basic considerations of
energetics and the broad-band spectra of heating sources, as well as
the specific model fits suggest that the nuclear starburst is the
dominant contributor to the heating of the dust in the central few
tens of parsecs of the galaxy.

More detailed modelling would show more
precisely the roles of the starburst and AGN in the heating of the
dust. For example, we have opted against including any
contribution heated by the central engine while fitting the starburst to avoid the 
near-IR problems inherent in the underlying RIAF spectrum. 
In this simplified model fitting, 
we have used only the N/Q ratio to determine the  model parameters.
A further complication that is beyond the scope of this
work would be to 
fit a broader spectral range and to include
luminosity contraints
simultaneously. 
Nonetheless, even the N/Q ratio by itself robustly excludes 
large portions of parameter space.
We emphasize that whatever its ultimate power source,
the mid-IR emission alone demonstrates that warm dust is present in
the nucleus of NGC~1097. The dust is local to the nucleus 
and must obscure it from certain lines of sight whilst leaving the broad lines unobscured
from our viewing angle, fulfilling at least two of the requirements of the AGN unified scheme.

\clearpage
\clearpage

\subsection{The spectra}
\label{spectrum}

\begin{figure}[t]
\includegraphics[angle=270,width=14cm]{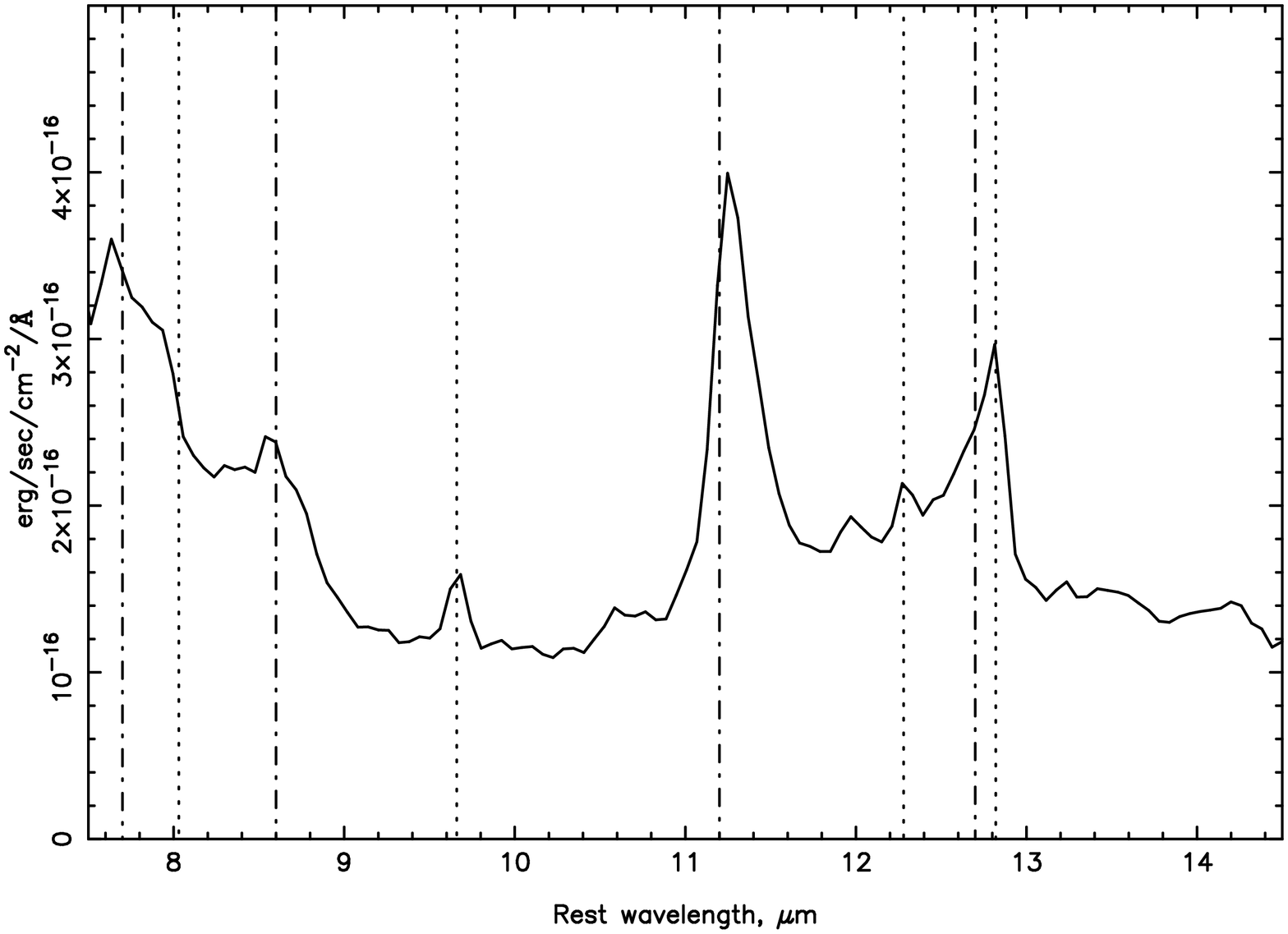}
\caption{Spitzer IRS spectrum of the nucleus NGC~1097. Lines are drawn at 7.7~\um\ (PAH), 8.03~\um\ (H$_{2}$ 0-0(S4)), 8.6~\um\ (PAH), 9.66~\um\ (H$_{2}$ 0-0(S3)), 11.2~\um\ (PAH), 12.28~\um\ (H$_{2}$ 0-0(S2)),12.7~\um\ (PAH) and 12.82~\um\ ([Ne II]). The peak wavelengths of the PAH bands vary somewhat from source to source \citep{vanDiedenhoven04}.}
\label{fig:IRS}
\end{figure}

The IRS short-low spectrum of NGC~1097 (Fig.~\ref{fig:IRS}) shows a number of prominent emission features, including the 7.7, 8.6, 11.2 and 12.7~\um\ PAH bands (this last line blended with the 12.82~\um\ [Ne II] fine structure line).  NGC~1097 is therefore one of the growing number of AGN in which strong PAH emission is detected within as little as 150 pc of the active nucleus.
\clearpage
\clearpage

To estimate the 3.3~\um\ band flux that would be expected from the 3.6 $\times$ 3.6\arcsec\ (approx. 300 $\times$ 300 pc) region covered by the IRS spectrum,  we use the finding of \citet{Hony01} that the 11.2~\um\ band is typically a factor of 3-4 stronger than the 3.3~\um\ feature in a wide range of environments. The flux in the 11.2~\um\ band, $7.1 \times 10^{-13}\rm erg~sec^{-1} cm^{-2}$ (as obtained by a spline fit to the continuum and direct integration under the band), then implies $\sim 2 \times 10^{-13}\rm erg~sec^{-1} cm^{-2}$  in the 3.3~\um\ feature. The corresponding luminosities are $2.5 \times 10^{40}\rm erg~sec^{-1} $ and $\sim 7 \times 10^{39}\rm erg~sec^{-1} $ for the 11.2 and 3.3~$\mu$m bands, respectively, within about a factor of two of the 3.3~$\mu$m PAH luminosity of Seyfert 1 galaxies such as NGC~4235 and MCG~-2-33-34 on similar physical scales \citep{Imanishi04}. 

\begin{figure}[t]
\includegraphics[angle=0,width=14cm]{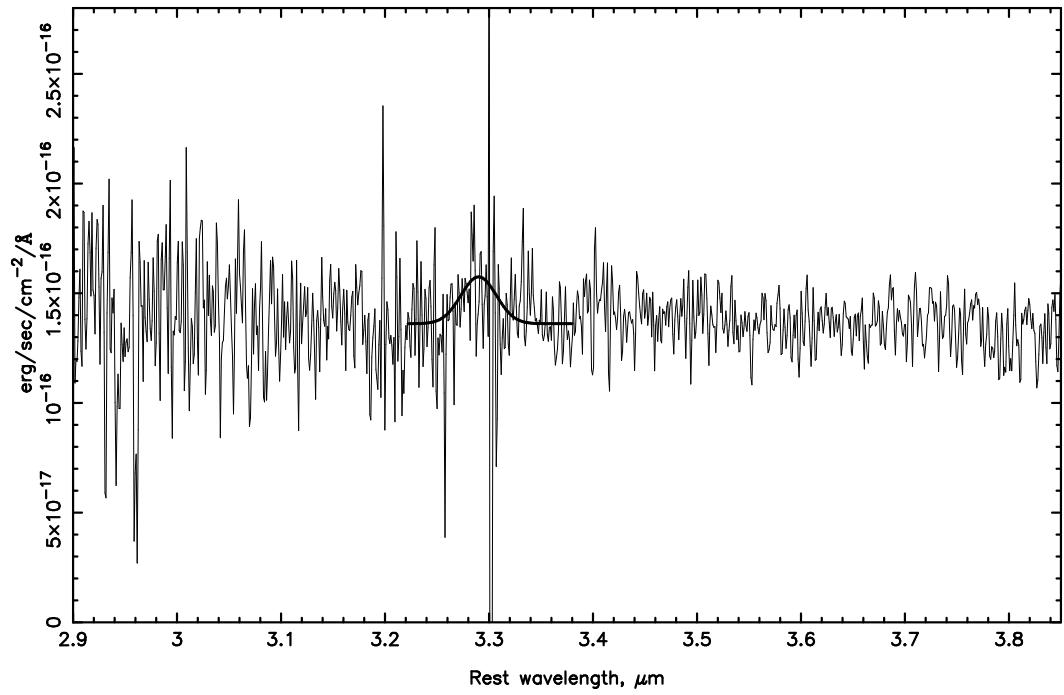}
\caption{L-band spectrum of NGC~1097. The black line represents the strongest  3.3~\um\ PAH band likely to be present in the spectrum (see text).}
\label{fig:GNIRS}
\end{figure}

In the GNIRS spectrum of the central 0.45 $\times$ 0.45\arcsec, however, no 3.3~$\mu$m feature is evident (Fig.~\ref{fig:GNIRS}). Using the 95\% confidence interval on the difference between the means of points near the peak of and just outside the feature \citep[and assuming a Gaussian profile with FWHM=0.04~$\mu$m;][]{vanDiedenhoven04}, we estimate the 3.3~$\mu$m PAH flux to be $\la 9 \times 10^{-15}\rm erg~sec^{-1} cm^{-2}$ . This is well below the flux that would be expected if the carriers of the 11.2~$\mu$m band were concentrated in the unresolved mid-IR source, as might be expected if they existed predominantly in warm, dense, dusty nuclear clouds, even allowing for uncertainties in the flux calibration of the GNIRS and IRS spectra. The spatial distribution of the mid-IR and PAH emission is very different.

\clearpage
\clearpage

Comparison of the N band and L' band continuum fluxes in similar apertures \citep{Telesco81,AlonsoHerrero01}, combined with the above 11.2/3.3~$\mu$m band ratio, suggests that 3.3~$\mu$m band emission would peak at nearly twice the continuum level in our spectrum if the PAH band- and continuum-emitting material were similarly distributed over the central region of NGC~1097.   That would imply an equivalent width (EW$_{3.3}$) $\sim$40 nm, whereas the maximum likely 3.3~$\mu$m feature flux estimated above implies  EW$_{3.3} \la$7 nm. This puts the EW$_{3.3}$ of the nuclear starburst in NGC~1097 at the low end of the values reported for few-hundred-parsec-scale starbursts by \citet{Imanishi03}, although comparable to several of those in \citet{Imanishi04}. The lack of PAH emission in the nucleus implies an increase in the relative level of featureless L-band continuum in this aperture, possibly due to destruction of the PAH molecules \citep[and/or dehydrogenation and ionization, which affect the molecules' ability to emit at 3.3~$\mu$m and their likelihood of photodissociation;][]{Allain96,Allamandola99}. This is examined in more detail in \S~\ref{PAHdiscuss}.


\section{Discussion}
\label{discuss}

\subsection{The nuclear mid-IR emission}
\label{MIRdiscuss}

The detection of UV variability, hard X-ray spectra and polarized broad emission lines in a number of LINERs \citep[e.g.][]{Barth99,Terashima03,Maoz05} has demonstrated that at least some of these objects harbor accreting supermassive black holes at their centers. 
An obvious question, then, is whether these LINERs possess the dusty tori postulated by the AGN unified scheme.

In the case of NGC~1097, the detection of unresolved thermal emission implies that there is warm dust in the nucleus of this galaxy that would obscure the nuclear broad lines if viewed from certain angles. As with Seyfert nuclei, the dust is compact (r$<$19 pc) at mid-IR wavelengths. On these two counts at least, it satisfies the requirements of the AGN unified model. On the other hand, our modeling shows that the accretion-powered nucleus of NGC~1097 is incapable of heating the dust efficiently enough to produce the observed mid-IR emission. The torus demanded by the unified model, if it exists at all, should be about a factor of two weaker in the mid-IR than the entity that we have detected, and have a different SED. The dust that we observe is not heated by the central engine and is therefore not necessarily directly associated with the AGN, consistent with the prediction of \citet{Elitzur06} that tori should disappear at $L<10^{42}\rm erg \; s^{-1}$. 

This does not entirely rule out, however, that NGC~1097 hosts a dusty torus.  If a cool, IR-faint torus previously existed, Fig.~\ref{fig:modelsb} shows that the young nuclear star cluster is capable of heating the dust so that it emits quite strongly in the mid-IR. Alternatively, the starburst itself could be the obscurer of the unified model. For this to be plausible, the dust associated with the starburst must provide only partial covering and star formation must be common in the innermost regions of at least low-luminosity AGN. The geometry of the dust is not well constrained by the observations and modeling presented here, but \citet{Prieto05} detect cold, dusty filaments spiraling down into the central 10 pc of NGC~1097, apparently in a thin disk geometry, and a kinematic study by \citet{Fathi06} shows material streaming into this central 10 pc region. \citet{Fathi06} suggest that the infalling material may have triggered the nuclear star formation, and disks of fairly recently-formed stars seem quite common in the central $\sim$50 pc of Seyfert galaxies \citep{Davies06a}. Obscuration by dusty starbursts seems at odds with the fact that low-luminosity AGN appear to have unobscured nuclei  \citep[about as many of the 17 LINERs with known nuclear UV point sources are type 2s as type 1s; ][]{Maoz05}, but could still apply to some fraction of the population. An observational assessment of the frequency of very small-scale star formation in these objects would be valuable in this context.

To the best of our knowledge, small-scale nuclear dust emission has been searched for in only a handful of other LINERs \citep{Chary00,Grossan01,Perlman01,Whysong04}, of which such emission was detected only in M81.  \citet{Grossan01} note that the SED of M81 resembles those of Seyfert galaxies and quasars, suggesting that the nuclear 10~$\mu$m emission in that object is also from a warm, central dust concentration. These results show that at least some LINERs do have dusty nuclei, but it is not yet clear whether these galaxies, particularly NGC~1097 with its nuclear starburst and infalling material, are unusual cases. High spatial resolution infrared observations of a larger sample of objects would help to establish whether or not this is true of AGN-containing LINERs in general, elucidate the mechanisms governing the presence or absence of the dust, and clarify how these low-luminosity objects may fit into unified schemes of AGN.

\subsection{The lack of strong PAH emission at the nucleus}
\label{PAHdiscuss}

As discussed in \S~\ref{spectrum}, 3.3~$\mu$m PAH emission is not detected in the nucleus of NGC~1097, although a strong 11.2~$\mu$m band is seen in the larger-aperture IRS spectrum (Figures~\ref{fig:IRS} \& \ref{fig:GNIRS}).  Most of the L-band flux in the smaller aperture must arise in material that does not produce a 3.3~$\mu$m band and whose emission becomes more important in smaller apertures. Likely sources of such emission include hot dust around the AGN and nuclear stellar cluster, and the central engine itself.  The intrinsic AGN/RIAF contribution to the L-band spectrum is not taken into account in Fig.~\ref{fig:modelsb}, but it would have to account for most of the total in order to "dilute" the PAH emission enough to avoid its detection in the GNIRS spectrum. The model fits suggest that this is unlikely (most of the L-band emission is thermal), even taking into account the simplicity of the starburst model, but some contribution is certainly possible. As for hot dust, that dust must be depleted (relative to the surrounding regions) in PAH molecules able to emit at 3.3~$\mu$m.

Both the AGN and starburst may contribute to suppressing the PAH emission.
Calculations suggest that PAH molecules should be destroyed close to luminous AGN; in particular, 
\citet{Sieb04}  estimate that the PAH band carriers should be destroyed by soft X-ray fluxes $\ga10^{3} \rm erg\;s^{-1} cm^{-2}$. Extending the spectral model of \citet{Nemmen06} and correcting
for extinction, the intrinsic AGN luminosity from 0.02 to 2 keV is $6.2\times10^{40} \rm erg\;s^{-1}$.
At a distance of 150 pc from the nucleus --- the most distant regions covered by the IRS spectrum --- 
the flux is then only about $10^{-2} \rm erg\;s^{-1}cm^{-2}$, consistent with the
appearance of the PAH bands in the {\em Spitzer} data. Even at only 1
pc, though, the (current) soft X-ray flux from the AGN is a factor of a few below the PAH destruction threshold.
 
While the starburst does not provide much X-ray emission, it
is luminous at UV wavelengths.  The $10^6 M_\sun$, $10^6$ year-old
starburst  has an intrinsic luminosity above 0.02 keV
of $2.1\times10^{42} \rm erg\;s^{-1}$ \citep{Leitherer99}.  
Because of the starburst's softer radiation field, 
the average flux must exceed $\sim10^{4} \rm erg\;s^{-1} cm^{-2}$ 
to destroy PAHs \citep{Sieb04}.
The starburst would be unlikely to destroy the distant
molecules detected in the IRS spectrum, but it could destroy the
PAHs located within $\sim$1 pc. 

Observations of other stellar clusters of similar mass and age support the idea that the stellar cluster itself is partly responsible for the lack of PAH emission in its immediate surroundings. \citet{Tacconi05} recently showed that, while the peak 3.3~$\mu$m band flux occurs in the general nuclear regions of the starburst galaxies NGC~253 and NGC~1808, the feature-to-continuum ratio actually reaches a minimum at the locations of maximum star formation activity in those nuclei. This is ascribed to photoionization or photodestruction of the PAH molecules in the radiation field of the stellar clusters. The strong PAH emission from the massive, young clusters in galaxies like He~2-10  \citep[][]{AlonsoHerrero04} might simply arise in diffuse material outside the immediate vicinity of the star clusters but still included in the 1.6\arcsec\ slit used for that work; a similar explanation  has been suggested for the PAH bands in the Antennae galaxies \citep{Snijders06}. This would account for the strength of the PAH bands in the {\em Spitzer} spectrum of NGC~1097 and their weakness at the nucleus.

To summarise, the 3.3~$\mu$m band has been a useful tracer of star formation on fairly small scales ($\sim$ 60 - few hundred pc) in AGN \citep{Imanishi02,Imanishi03,Rodriguez03,Imanishi04}. However, if NGC~1097 turns out to be typical, the combined effects of AGN continuum and the radiation field of stellar clusters themselves probably mean that it is unlikely that the 3.3~$\mu$m feature will be as useful a diagnostic of vigorous star formation occurring on the smallest (few parsec) scales.


\section{Conclusions}
\label{conc}

We have performed mid-IR imaging and L-band spectroscopy of the central region of NGC~1097, at $\la$0.5\arcsec\ angular resolution, to search for thermal dust emission from the active nucleus and to test a potential diagnostic of small-scale nuclear star formation. A compact nuclear source is detected at both 11.7 and 18.3~$\mu$m, with a luminosity well in excess of that predicted by the accretion disk + RIAF + jet model of \citet{Nemmen06}. This mid-IR ``excess'' is strong evidence for thermal emission from dust in the nucleus of this low-luminosity AGN. 

Fitting clumpy obscuration models to this and other high-resolution IR data suggests that the star cluster detected by \citet{StorchiBergmann05}, not the central engine,  is the dominant contributor to the heating of the dust in the nucleus. In fact, tori heated solely by low-luminosity AGN --- if they exist at all --- are likely to be relatively faint even in the mid-IR close to where their thermal emission peaks. This may mean that the mid-IR emission that we detect is not from the torus of AGN unified schemes, in accordance with certain predictions that tori should not exist in AGN with luminosities as low as that of NGC~1097. Alternatively it is possible, although not required by the data, that dust associated with the small-scale nuclear starburst is itself the obscuring material of the unified model. Further observations of IR emission and signatures of nuclear star formation in low-luminosity AGN would help distinguish between these possibilities.

Despite the likely energetic importance of the starburst, the 3.3~$\mu$m PAH band is not detected in the central 19 pc of this galaxy, although strong 11.2~$\mu$m PAH emission is observed in the surrounding areas.  Destruction and/or ionization of the PAH molecules in the radiation field of the stellar cluster itself are likely to account for the non-detection of the 3.3~$\mu$m band. These factors may limit the use of PAH features as extinction-insensitive tracers of very compact nuclear starbursts associated with dusty tori in other low-luminosity AGN.

\acknowledgments

We would like to thank A. Alonso-Herrero, M. Eracleous, T. Geballe  and M. A. Prieto for providing helpful information, and the anonymous referee for a careful reading of the manuscript. NAL acknowledges work supported by the National Science Foundation
under Grant No. 0237291.
CP acknowledges work supported by the National Science Foundation under Grant    
No. 0206617. This paper is based on observations obtained at the Gemini Observatory (Program ID: GS-2005B-Q-29),
which is operated by the Association of Universities for Research in
Astronomy, Inc., under a cooperative agreement with the NSF on behalf of
the Gemini partnership: the National Science Foundation (United States),
the Particle Physics and Astronomy Research Council (United Kingdom), the
National Research Council (Canada), CONICYT (Chile), the Australian
Research Council (Australia), CNPq (Brazil) and CONICET (Argentina). This work is also based in part on observations made with the Spitzer Space Telescope, which is operated by the Jet Propulsion Laboratory, California Institute of Technology under a contract with NASA.


\begin{thebibliography}{73}
\expandafter\ifx\csname natexlab\endcsname\relax\def\natexlab#1{#1}\fi

\bibitem[{{Allain} {et~al.}(1996){Allain}, {Leach}, \& {Sedlmayr}}]{Allain96}
{Allain}, T., {Leach}, S., \& {Sedlmayr}, E. 1996, \aap, 305, 616

\bibitem[{{Allamandola} {et~al.}(1999){Allamandola}, {Hudgins}, \&
  {Sandford}}]{Allamandola99}
{Allamandola}, L.~J., {Hudgins}, D.~M., \& {Sandford}, S.~A. 1999, \apjl, 511,
  L115

\bibitem[{{Alonso-Herrero} {et~al.}(2001){Alonso-Herrero}, {Engelbracht},
  {Rieke}, {Rieke}, \& {Quillen}}]{AlonsoHerrero01}
{Alonso-Herrero}, A., {Engelbracht}, C.~W., {Rieke}, M.~J., {Rieke}, G.~H., \&
  {Quillen}, A.~C. 2001, ApJ, 546, 952

\bibitem[{{Alonso-Herrero} {et~al.}(2004){Alonso-Herrero}, {Takagi}, {Baker},
  {Rieke}, {Rieke}, {Imanishi}, \& {Scoville}}]{AlonsoHerrero04}
{Alonso-Herrero}, A., {Takagi}, T., {Baker}, A.~J., {Rieke}, G.~H., {Rieke},
  M.~J., {Imanishi}, M., \& {Scoville}, N.~Z. 2004, \apj, 612, 222

\bibitem[{{Antonucci} \& {Miller}(1985)}]{AM85}
{Antonucci}, R.~R.~J., \& {Miller}, J.~S. 1985, ApJ, 297, 621

\bibitem[{{Barth} {et~al.}(1999){Barth}, {Filippenko}, \& {Moran}}]{Barth99}
{Barth}, A.~J., {Filippenko}, A.~V., \& {Moran}, E.~C. 1999, \apj, 525, 673

\bibitem[{{Bock} {et~al.}(2000){Bock}, {Neugebauer}, {Matthews}, {Soifer},
  {Becklin}, {Ressler}, {Marsh}, {Werner}, {Egami}, \& {Blandford}}]{Bock2}
{Bock}, J.~J., {Neugebauer}, G., {Matthews}, K., {Soifer}, B.~T., {Becklin},
  E.~E., {Ressler}, M., {Marsh}, K., {Werner}, M.~W., {Egami}, E., \&
  {Blandford}, R. 2000, AJ, 120, 2904

\bibitem[{{Chary} {et~al.}(2000){Chary}, {Becklin}, {Evans}, {Neugebauer},
  {Scoville}, {Matthews}, \& {Ressler}}]{Chary00}
{Chary}, R., {Becklin}, E.~E., {Evans}, A.~S., {Neugebauer}, G., {Scoville},
  N.~Z., {Matthews}, K., \& {Ressler}, M.~E. 2000, \apj, 531, 756

\bibitem[{{Cid Fernandes} \& {Terlevich}(1995)}]{CidFernandes95}
{Cid Fernandes}, R.~J., \& {Terlevich}, R. 1995, \mnras, 272, 423

\bibitem[{{Colina} {et~al.}(2002){Colina}, {Gonzalez Delgado}, {Mas-Hesse}, \&
  {Leitherer}}]{Colina02}
{Colina}, L., {Gonzalez Delgado}, R., {Mas-Hesse}, J.~M., \& {Leitherer}, C.
  2002, \apj, 579, 545

\bibitem[{{Davies} {et~al.}(2006{\natexlab{a}}){Davies}, {Genzel}, {Tacconi},
  {Mueller Sanchez}, \& {Sternberg}}]{Davies06a}
{Davies}, R., {Genzel}, R., {Tacconi}, L., {Mueller Sanchez}, F., \&
  {Sternberg}, A. 2006{\natexlab{a}}, ArXiv Astrophysics e-prints

\bibitem[{{Davies} {et~al.}(2006{\natexlab{b}}){Davies}, {Thomas}, {Genzel},
  {Mueller S{\'a}nchez}, {Tacconi}, {Sternberg}, {Eisenhauer}, {Abuter},
  {Saglia}, \& {Bender}}]{Davies06}
{Davies}, R.~I., {Thomas}, J., {Genzel}, R., {Mueller S{\'a}nchez}, F.,
  {Tacconi}, L.~J., {Sternberg}, A., {Eisenhauer}, F., {Abuter}, R., {Saglia},
  R., \& {Bender}, R. 2006{\natexlab{b}}, \apj, 646, 754

\bibitem[{{Elitzur} \& {Shlosman}(2006)}]{Elitzur06}
{Elitzur}, M., \& {Shlosman}, I. 2006, \apjl, 648, L101

\bibitem[{{Fabian} {et~al.}(1998){Fabian}, {Barcons}, {Almaini}, \&
  {Iwasawa}}]{Fabian98}
{Fabian}, A.~C., {Barcons}, X., {Almaini}, O., \& {Iwasawa}, K. 1998, \mnras,
  297, L11+

\bibitem[{{Fathi} {et~al.}(2006){Fathi}, {Storchi-Bergmann}, {Riffel}, {Winge},
  {Axon}, {Robinson}, {Capetti}, \& {Marconi}}]{Fathi06}
{Fathi}, K., {Storchi-Bergmann}, T., {Riffel}, R.~A., {Winge}, C., {Axon},
  D.~J., {Robinson}, A., {Capetti}, A., \& {Marconi}, A. 2006, \apjl, 641, L25

\bibitem[{{Genzel} {et~al.}(1998){Genzel}, {Lutz}, {Sturm}, {Egami}, {Kunze},
  {Moorwood}, {Rigopoulou}, {Spoon}, {Sternberg}, {Tacconi-Garman}, {Tacconi},
  \& {Thatte}}]{Genzel98}
{Genzel}, R., {Lutz}, D., {Sturm}, E., {Egami}, E., {Kunze}, D., {Moorwood},
  A.~F.~M., {Rigopoulou}, D., {Spoon}, H.~W.~W., {Sternberg}, A.,
  {Tacconi-Garman}, L.~E., {Tacconi}, L., \& {Thatte}, N. 1998, \apj, 498, 579

\bibitem[{{Gonz{\'a}lez Delgado} {et~al.}(1998){Gonz{\'a}lez Delgado},
  {Heckman}, {Leitherer}, {Meurer}, {Krolik}, {Wilson}, {Kinney}, \&
  {Koratkar}}]{Gonzalez98}
{Gonz{\'a}lez Delgado}, R.~M., {Heckman}, T., {Leitherer}, C., {Meurer}, G.,
  {Krolik}, J., {Wilson}, A.~S., {Kinney}, A., \& {Koratkar}, A. 1998, \apj,
  505, 174

\bibitem[{{Gorjian} {et~al.}(2004){Gorjian}, {Werner}, {Jarrett}, {Cole}, \&
  {Ressler}}]{Gorjian04}
{Gorjian}, V., {Werner}, M.~W., {Jarrett}, T.~H., {Cole}, D.~M., \& {Ressler},
  M.~E. 2004, \apj, 605, 156

\bibitem[{{Grossan} {et~al.}(2001){Grossan}, {Gorjian}, {Werner}, \&
  {Ressler}}]{Grossan01}
{Grossan}, B., {Gorjian}, V., {Werner}, M., \& {Ressler}, M. 2001, \apj, 563,
  687

\bibitem[{{Hony} {et~al.}(2001){Hony}, {Van Kerckhoven}, {Peeters}, {Tielens},
  {Hudgins}, \& {Allamandola}}]{Hony01}
{Hony}, S., {Van Kerckhoven}, C., {Peeters}, E., {Tielens}, A.~G.~G.~M.,
  {Hudgins}, D.~M., \& {Allamandola}, L.~J. 2001, \aap, 370, 1030

\bibitem[{{Houck} {et~al.}(2004){Houck}, {Roellig}, {van Cleve}, {Forrest},
  {Herter}, {Lawrence}, {Matthews}, {Reitsema}, {Soifer}, {Watson}, {Weedman},
  {Huisjen}, {Troeltzsch}, {Barry}, {Bernard-Salas}, {Blacken}, {Brandl},
  {Charmandaris}, {Devost}, {Gull}, {Hall}, {Henderson}, {Higdon}, {Pirger},
  {Schoenwald}, {Sloan}, {Uchida}, {Appleton}, {Armus}, {Burgdorf},
  {Fajardo-Acosta}, {Grillmair}, {Ingalls}, {Morris}, \& {Teplitz}}]{Houck04}
{Houck}, J.~R., {Roellig}, T.~L., {van Cleve}, J., {Forrest}, W.~J., {Herter},
  T., {Lawrence}, C.~R., {Matthews}, K., {Reitsema}, H.~J., {Soifer}, B.~T.,
  {Watson}, D.~M., {Weedman}, D., {Huisjen}, M., {Troeltzsch}, J., {Barry},
  D.~J., {Bernard-Salas}, J., {Blacken}, C.~E., {Brandl}, B.~R.,
  {Charmandaris}, V., {Devost}, D., {Gull}, G.~E., {Hall}, P., {Henderson},
  C.~P., {Higdon}, S.~J.~U., {Pirger}, B.~E., {Schoenwald}, J., {Sloan}, G.~C.,
  {Uchida}, K.~I., {Appleton}, P.~N., {Armus}, L., {Burgdorf}, M.~J.,
  {Fajardo-Acosta}, S.~B., {Grillmair}, C.~J., {Ingalls}, J.~G., {Morris},
  P.~W., \& {Teplitz}, H.~I. 2004, \apjs, 154, 18

\bibitem[{{Imanishi}(2002)}]{Imanishi02}
{Imanishi}, M. 2002, ApJ, 569, 44

\bibitem[{{Imanishi}(2003)}]{Imanishi03}
---. 2003, \apj, 599, 918

\bibitem[{{Imanishi} \& {Wada}(2004)}]{Imanishi04}
{Imanishi}, M., \& {Wada}, K. 2004, \apj, 617, 214

\bibitem[{{Jaffe} {et~al.}(2004){Jaffe}, {Meisenheimer}, {R{\" o}ttgering},
  {Leinert}, {Richichi}, {Chesneau}, {Fraix-Burnet}, {Glazenborg-Kluttig},
  {Granato}, {Graser}, {Heijligers}, {K{\" o}hler}, {Malbet}, {Miley},
  {Paresce}, {Pel}, {Perrin}, {Przygodda}, {Schoeller}, {Sol}, {Waters},
  {Weigelt}, {Woillez}, \& {de Zeeuw}}]{Jaffe}
{Jaffe}, W., {Meisenheimer}, K., {R{\" o}ttgering}, H.~J.~A., {Leinert}, C.,
  {Richichi}, A., {Chesneau}, O., {Fraix-Burnet}, D., {Glazenborg-Kluttig}, A.,
  {Granato}, G.-L., {Graser}, U., {Heijligers}, B., {K{\" o}hler}, R.,
  {Malbet}, F., {Miley}, G.~K., {Paresce}, F., {Pel}, J.-W., {Perrin}, G.,
  {Przygodda}, F., {Schoeller}, M., {Sol}, H., {Waters}, L.~B.~F.~M.,
  {Weigelt}, G., {Woillez}, J., \& {de Zeeuw}, P.~T. 2004, Nature, 429, 47

\bibitem[{{Laurent} {et~al.}(2000){Laurent}, {Mirabel}, {Charmandaris},
  {Gallais}, {Madden}, {Sauvage}, {Vigroux}, \& {Cesarsky}}]{Laurent00}
{Laurent}, O., {Mirabel}, I.~F., {Charmandaris}, V., {Gallais}, P., {Madden},
  S.~C., {Sauvage}, M., {Vigroux}, L., \& {Cesarsky}, C. 2000, \aap, 359, 887

\bibitem[{{Le Floc'h} {et~al.}(2001){Le Floc'h}, {Mirabel}, {Laurent},
  {Charmandaris}, {Gallais}, {Sauvage}, {Vigroux}, \& {Cesarsky}}]{Lefloch}
{Le Floc'h}, E., {Mirabel}, I.~F., {Laurent}, O., {Charmandaris}, V.,
  {Gallais}, P., {Sauvage}, M., {Vigroux}, L., \& {Cesarsky}, C. 2001, A\&A,
  367, 487

\bibitem[{{Leitherer} {et~al.}(1999){Leitherer}, {Schaerer}, {Goldader},
  {Delgado}, {Robert}, {Kune}, {de Mello}, {Devost}, \&
  {Heckman}}]{Leitherer99}
{Leitherer}, C., {Schaerer}, D., {Goldader}, J.~D., {Delgado}, R.~M.~G.,
  {Robert}, C., {Kune}, D.~F., {de Mello}, D.~F., {Devost}, D., \& {Heckman},
  T.~M. 1999, \apjs, 123, 3

\bibitem[{{Lord}(1992)}]{Lord92}
{Lord}, S.~D. 1992, NASA Technical Memor. 103957

\bibitem[{{Lutz} {et~al.}(1998){Lutz}, {Spoon}, {Rigopoulou}, {Moorwood}, \&
  {Genzel}}]{Lutz98}
{Lutz}, D., {Spoon}, H.~W.~W., {Rigopoulou}, D., {Moorwood}, A.~F.~M., \&
  {Genzel}, R. 1998, \apjl, 505, L103

\bibitem[{{Maoz} {et~al.}(2005){Maoz}, {Nagar}, {Falcke}, \& {Wilson}}]{Maoz05}
{Maoz}, D., {Nagar}, N.~M., {Falcke}, H., \& {Wilson}, A.~S. 2005, \apj, 625,
  699

\bibitem[{{Mason} {et~al.}(2006){Mason}, {Geballe}, {Packham}, {Levenson},
  {Elitzur}, {Fisher}, \& {Perlman}}]{Mason06}
{Mason}, R.~E., {Geballe}, T.~R., {Packham}, C., {Levenson}, N.~A., {Elitzur},
  M., {Fisher}, R.~S., \& {Perlman}, E. 2006, \apj, 640, 612

\bibitem[{{Miller} \& {Goodrich}(1990)}]{Miller90}
{Miller}, J.~S., \& {Goodrich}, R.~W. 1990, \apj, 355, 456

\bibitem[{{Moorwood}(1986)}]{Moorwood86}
{Moorwood}, A.~F.~M. 1986, \aap, 166, 4

\bibitem[{{Moran} {et~al.}(2000){Moran}, {Barth}, {Kay}, \&
  {Filippenko}}]{Moran00}
{Moran}, E.~C., {Barth}, A.~J., {Kay}, L.~E., \& {Filippenko}, A.~V. 2000,
  \apjl, 540, L73

\bibitem[{{Mueller S{\'a}nchez} {et~al.}(2006){Mueller S{\'a}nchez}, {Davies},
  {Eisenhauer}, {Tacconi}, {Genzel}, \& {Sternberg}}]{MuellerSanchez06}
{Mueller S{\'a}nchez}, F., {Davies}, R.~I., {Eisenhauer}, F., {Tacconi}, L.~J.,
  {Genzel}, R., \& {Sternberg}, A. 2006, \aap, 454, 481

\bibitem[{{Nemmen} {et~al.}(2006){Nemmen}, {Storchi-Bergmann}, {Yuan},
  {Eracleous}, {Terashima}, \& {Wilson}}]{Nemmen06}
{Nemmen}, R.~S., {Storchi-Bergmann}, T., {Yuan}, F., {Eracleous}, M.,
  {Terashima}, Y., \& {Wilson}, A.~S. 2006, \apj, 643, 652

\bibitem[{{Nenkova} {et~al.}(2002){Nenkova}, {Ivezi{\' c}}, \&
  {Elitzur}}]{Nenkova}
{Nenkova}, M., {Ivezi{\' c}}, {\v Z}., \& {Elitzur}, M. 2002, ApJL, 570, L9

\bibitem[{{Packham} {et~al.}(2005){Packham}, {Radomski}, {Roche}, {Aitken},
  {Perlman}, {Alonso-Herrero}, {Colina}, \& {Telesco}}]{Packham05}
{Packham}, C., {Radomski}, J.~T., {Roche}, P.~F., {Aitken}, D.~K., {Perlman},
  E., {Alonso-Herrero}, A., {Colina}, L., \& {Telesco}, C.~M. 2005, ApJL, 618,
  L17

\bibitem[{{Peeters} {et~al.}(2004){Peeters}, {Spoon}, \& {Tielens}}]{Peeters04}
{Peeters}, E., {Spoon}, H.~W.~W., \& {Tielens}, A.~G.~G.~M. 2004, \apj, 613,
  986

\bibitem[{{Perlman} {et~al.}(2001){Perlman}, {Sparks}, {Radomski}, {Packham},
  {Fisher}, {Pi{\~n}a}, \& {Biretta}}]{Perlman01}
{Perlman}, E.~S., {Sparks}, W.~B., {Radomski}, J., {Packham}, C., {Fisher},
  R.~S., {Pi{\~n}a}, R., \& {Biretta}, J.~A. 2001, \apjl, 561, L51

\bibitem[{{Phillips} {et~al.}(1984){Phillips}, {Pagel}, {Edmunds}, \&
  {Diaz}}]{Phillips84}
{Phillips}, M.~M., {Pagel}, B.~E.~J., {Edmunds}, M.~G., \& {Diaz}, A. 1984,
  \mnras, 210, 701

\bibitem[{{Prieto} {et~al.}(2005){Prieto}, {Maciejewski}, \&
  {Reunanen}}]{Prieto05}
{Prieto}, M.~A., {Maciejewski}, W., \& {Reunanen}, J. 2005, \aj, 130, 1472

\bibitem[{{Prieto} {et~al.}(2004){Prieto}, {Meisenheimer}, {Marco}, {Reunanen},
  {Contini}, {Clenet}, {Davies}, {Gratadour}, {Henning}, {Klaas}, {Kotilainen},
  {Leinert}, {Lutz}, {Rouan}, \& {Thatte}}]{Prieto04b}
{Prieto}, M.~A., {Meisenheimer}, K., {Marco}, O., {Reunanen}, J., {Contini},
  M., {Clenet}, Y., {Davies}, R.~I., {Gratadour}, D., {Henning}, T., {Klaas},
  U., {Kotilainen}, J., {Leinert}, C., {Lutz}, D., {Rouan}, D., \& {Thatte}, N.
  2004, \apj, 614, 135

\bibitem[{{Radomski} {et~al.}(2003){Radomski}, {Pi{\~ n}a}, {Packham},
  {Telesco}, {De Buizer}, {Fisher}, \& {Robinson}}]{Radomski03}
{Radomski}, J.~T., {Pi{\~ n}a}, R.~K., {Packham}, C., {Telesco}, C.~M., {De
  Buizer}, J.~M., {Fisher}, R.~S., \& {Robinson}, A. 2003, \apj, 587, 117

\bibitem[{{Rigopoulou} {et~al.}(1999){Rigopoulou}, {Spoon}, {Genzel}, {Lutz},
  {Moorwood}, \& {Tran}}]{Rigopoulou99}
{Rigopoulou}, D., {Spoon}, H.~W.~W., {Genzel}, R., {Lutz}, D., {Moorwood},
  A.~F.~M., \& {Tran}, Q.~D. 1999, AJ, 118, 2625

\bibitem[{{Risaliti} {et~al.}(2006){Risaliti}, {Maiolino}, {Marconi}, {Sani},
  {Berta}, {Braito}, {Ceca}, {Franceschini}, \& {Salvati}}]{Risaliti06}
{Risaliti}, G., {Maiolino}, R., {Marconi}, A., {Sani}, E., {Berta}, S.,
  {Braito}, V., {Ceca}, R.~D., {Franceschini}, A., \& {Salvati}, M. 2006,
  \mnras, 365, 303

\bibitem[{{Roche} \& {Aitken}(1985)}]{Roche85}
{Roche}, P.~F., \& {Aitken}, D.~K. 1985, MNRAS, 215, 425

\bibitem[{{Roche} {et~al.}(1991){Roche}, {Aitken}, {Smith}, \&
  {Ward}}]{Roche91}
{Roche}, P.~F., {Aitken}, D.~K., {Smith}, C.~H., \& {Ward}, M.~J. 1991, MNRAS,
  248, 606

\bibitem[{{Roche} {et~al.}(2006{\natexlab{a}}){Roche}, {Packham}, {Aitken}, \&
  {Mason}}]{Roche06a}
{Roche}, P.~F., {Packham}, C., {Aitken}, D.~K., \& {Mason}, R.~E.
  2006{\natexlab{a}}, ArXiv Astrophysics e-prints

\bibitem[{{Roche} {et~al.}(2006{\natexlab{b}}){Roche}, {Packham}, {Telesco},
  {Radomski}, {Alonso-Hererro}, {Aitken}, {Colina}, \& {Perlman}}]{Roche06b}
{Roche}, P.~F., {Packham}, C., {Telesco}, C.~M., {Radomski}, J.~T.,
  {Alonso-Hererro}, A., {Aitken}, D.~K., {Colina}, L., \& {Perlman}, E.
  2006{\natexlab{b}}, \mnras, 258

\bibitem[{{Rodr{\'{\i}}guez-Ardila} \& {Viegas}(2003)}]{Rodriguez03}
{Rodr{\'{\i}}guez-Ardila}, A., \& {Viegas}, S.~M. 2003, \mnras, 340, L33

\bibitem[{{Sako} {et~al.}(2003){Sako}, {Okamoto}, {Kataza}, {Miyata}, {Takubo},
  {Honda}, {Fujiyoshi}, {Onaka}, \& {Yamashita}}]{Sako03}
{Sako}, S., {Okamoto}, Y.~K., {Kataza}, H., {Miyata}, T., {Takubo}, S.,
  {Honda}, M., {Fujiyoshi}, T., {Onaka}, T., \& {Yamashita}, T. 2003, \pasp,
  115, 1407

\bibitem[{{Siebenmorgen} {et~al.}(2004){Siebenmorgen}, {Kr{\" u}gel}, \&
  {Spoon}}]{Sieb04}
{Siebenmorgen}, R., {Kr{\" u}gel}, E., \& {Spoon}, H.~W.~W. 2004, A\&A, 414,
  123

\bibitem[{{Snijders} {et~al.}(2006){Snijders}, {van der Werf}, {Brandl},
  {Mengel}, {Schaerer}, \& {Wang}}]{Snijders06}
{Snijders}, L., {van der Werf}, P.~P., {Brandl}, B.~R., {Mengel}, S.,
  {Schaerer}, D., \& {Wang}, Z. 2006, \apjl, 648, L25

\bibitem[{{Soifer} {et~al.}(2003){Soifer}, {Bock}, {Marsh}, {Neugebauer},
  {Matthews}, {Egami}, \& {Armus}}]{Soifer03}
{Soifer}, B.~T., {Bock}, J.~J., {Marsh}, K., {Neugebauer}, G., {Matthews}, K.,
  {Egami}, E., \& {Armus}, L. 2003, \aj, 126, 143

\bibitem[{{Storchi-Bergmann} {et~al.}(1993){Storchi-Bergmann}, {Baldwin}, \&
  {Wilson}}]{StorchiBergmann93}
{Storchi-Bergmann}, T., {Baldwin}, J.~A., \& {Wilson}, A.~S. 1993, \apjl, 410,
  L11

\bibitem[{{Storchi-Bergmann} {et~al.}(2005){Storchi-Bergmann}, {Nemmen},
  {Spinelli}, {Eracleous}, {Wilson}, {Filippenko}, \&
  {Livio}}]{StorchiBergmann05}
{Storchi-Bergmann}, T., {Nemmen}, R.~S., {Spinelli}, P.~F., {Eracleous}, M.,
  {Wilson}, A.~S., {Filippenko}, A.~V., \& {Livio}, M. 2005, \apjl, 624, L13

\bibitem[{{Storchi-Bergmann} {et~al.}(2003){Storchi-Bergmann}, {Nemmen da
  Silva}, {Eracleous}, {Halpern}, {Wilson}, {Filippenko}, {Ruiz}, {Smith}, \&
  {Nagar}}]{StorchiBergmann03}
{Storchi-Bergmann}, T., {Nemmen da Silva}, R., {Eracleous}, M., {Halpern},
  J.~P., {Wilson}, A.~S., {Filippenko}, A.~V., {Ruiz}, M.~T., {Smith}, R.~C.,
  \& {Nagar}, N.~M. 2003, \apj, 598, 956

\bibitem[{{Storchi-Bergmann} {et~al.}(2000){Storchi-Bergmann}, {Raimann},
  {Bica}, \& {Fraquelli}}]{StorchiBergmann01}
{Storchi-Bergmann}, T., {Raimann}, D., {Bica}, E.~L.~D., \& {Fraquelli}, H.~A.
  2000, \apj, 544, 747

\bibitem[{{Tacconi-Garman} {et~al.}(2005){Tacconi-Garman}, {Sturm}, {Lehnert},
  {Lutz}, {Davies}, \& {Moorwood}}]{Tacconi05}
{Tacconi-Garman}, L.~E., {Sturm}, E., {Lehnert}, M., {Lutz}, D., {Davies},
  R.~I., \& {Moorwood}, A.~F.~M. 2005, \aap, 432, 91

\bibitem[{{Telesco} \& {Gatley}(1981)}]{Telesco81}
{Telesco}, C.~M., \& {Gatley}, I. 1981, \apjl, 247, L11

\bibitem[{{Telesco} {et~al.}(1998){Telesco}, {Pina}, {Hanna}, {Julian}, {Hon},
  \& {Kisko}}]{Telesco98}
{Telesco}, C.~M., {Pina}, R.~K., {Hanna}, K.~T., {Julian}, J.~A., {Hon}, D.~B.,
  \& {Kisko}, T.~M. 1998, in Proc. SPIE Vol. 3354, p. 534-544, Infrared
  Astronomical Instrumentation, Albert M. Fowler; Ed., ed. A.~M. {Fowler},
  534--544

\bibitem[{{Terashima} {et~al.}(2002){Terashima}, {Iyomoto}, {Ho}, \&
  {Ptak}}]{Terashima02}
{Terashima}, Y., {Iyomoto}, N., {Ho}, L.~C., \& {Ptak}, A.~F. 2002, \apjs, 139,
  1

\bibitem[{{Terashima} \& {Wilson}(2003)}]{Terashima03}
{Terashima}, Y., \& {Wilson}, A.~S. 2003, \apj, 583, 145

\bibitem[{{Tran}(2001)}]{Tran01}
{Tran}, H.~D. 2001, \apjl, 554, L19

\bibitem[{{van Diedenhoven} {et~al.}(2004){van Diedenhoven}, {Peeters}, {Van
  Kerckhoven}, {Hony}, {Hudgins}, {Allamandola}, \&
  {Tielens}}]{vanDiedenhoven04}
{van Diedenhoven}, B., {Peeters}, E., {Van Kerckhoven}, C., {Hony}, S.,
  {Hudgins}, D.~M., {Allamandola}, L.~J., \& {Tielens}, A.~G.~G.~M. 2004, \apj,
  611, 928

\bibitem[{{Voit}(1992)}]{Voit92}
{Voit}, G.~M. 1992, \mnras, 258, 841

\bibitem[{{Wada} \& {Norman}(2002)}]{Wada02}
{Wada}, K., \& {Norman}, C.~A. 2002, \apjl, 566, L21

\bibitem[{{Watabe} \& {Umemura}(2005)}]{Watabe05}
{Watabe}, Y., \& {Umemura}, M. 2005, \apj, 618, 649

\bibitem[{{Whysong} \& {Antonucci}(2004)}]{Whysong04}
{Whysong}, D., \& {Antonucci}, R. 2004, \apj, 602, 116

\bibitem[{{Young} {et~al.}(1996){Young}, {Hough}, {Efstathiou}, {Wills},
  {Bailey}, {Ward}, \& {Axon}}]{Young96b}
{Young}, S., {Hough}, J.~H., {Efstathiou}, A., {Wills}, B.~J., {Bailey}, J.~A.,
  {Ward}, M.~J., \& {Axon}, D.~J. 1996, MNRAS, 281, 1206

\bibitem[{{Zier} \& {Biermann}(2002)}]{Zier02}
{Zier}, C., \& {Biermann}, P.~L. 2002, \aap, 396, 91

\end{thebibliography}


\end{document}